\documentclass[epj-spec]{svjour}
\usepackage{graphics}
\usepackage{epsf,epsfig}

\newcommand{\ci}[1]{\cite{#1}}
\def\xb{\overline{x}}

\def\als{\alpha_s}
\def\vk{{\bf k}_{\perp}}
\def\vbs{{\bf b}}
\def\gev{\,{\rm GeV}}
\begin{document}
\title{GPDs and spin effects in light  mesons production.}
\author{S.V.Goloskokov \thanks{\email{goloskkv@theor.jinr.ru}} }
\institute{ Bogoliubov Laboratory of Theoretical  Physics,
  Joint Institute for Nuclear Research, Dubna, Russia}
\abstract{Light  meson electroproduction  is analyzed on the basis
of the handbag approach. Our results on the cross section and spin
effects are in good agrement with experiments at HERA, COMPASS and
HERMES energies. Predictions for $A_{UT}$ asymmetry for various
reactions are presented.
} 
\maketitle

 \section{Introduction}
In this report, we investigate light meson electroproduction  at
large photon virtualities $Q^2$ which gives access to Generalized
parton distributions (GPDs). Really, at small $x$-Bjorken  the
leading twist amplitude with longitudinally polarized photon and
vector meson (LL amplitude) at large $Q^2$ factorizes \cite{fact}
in the handbag approach into hard meson electroproduction off
partons and GPDs. In our calculations \cite{gk05,gk06,gk07q}, we
modify this approach by including the quark transverse degrees of
freedom and the Sudakov corrections in the hard subprocess
amplitude. Taking into account  the quark transverse momentum in
this modified perturbative approach (MPA) gives a possibility to
regularize the end-point singularities in the higher twist TT
amplitude which is essential in description of spin effects.  GPDs
are modeled with the help of double distribution representation.
Within the MPA we calculate the LL and TT amplitudes and,
subsequently, the cross sections and the spin observables in the
light meson electroproduction.  Modeling of the GPDs $E$ permits
us to extend our analysis to a transversally polarized target. We
calculate cross sections and the $A_{UT}$ asymmetry for various
vector meson production \cite{gk08}. The new results to the pion
electroproduction \cite{gk09}  presented here give access to the
GPDs $\tilde H$ and $\tilde E$.

Our results are in good agreement  with high energy  HERA
\cite{h1,zeus} experiments  and low energy COMPASS \cite{compass}
and HERMES \cite{hermes} data on meson electroproducion at small
and moderate $x$ \cite{gk05,gk06,gk07q,gk08,gk09}.

\section{GPDs and electroproduction of  vector mesons in the handbag approach}

In the model, the amplitude of the vector meson production off the
proton  reads as a convolution of the hard partonic subprocess
 ${\cal H}^V$ and $\hat H_i$
\begin{eqnarray}\label{amptt-nf-ji}
  {\cal M}^{Vi}_{\mu'\pm,\mu +} &=& \frac{e}{2}\, \sum_{a}e_a\,{\cal
  C}_a^{V}\, \sum_{\lambda}
         \int_{xi}^1 d\xb
        {\cal H}^{Vi}_{\mu'\lambda,\mu \lambda}
                                   \hat H_i(\xb,\xi,t) ,
\end{eqnarray}
where  $i$ denotes the gluon and quark contribution, sum over $a$
includes quark flavor $a$ and $C_a^{V}$ are the corresponding
flavor factors \cite{gk06};
 $\mu$ ($\mu'$) is the helicity of the photon (meson), and $\xb$
 is the momentum fraction of the parton with helicity $\lambda$.

 The $\hat H_i$ in (\ref{amptt-nf-ji}) are expressed in terms of GPDs.
 For the proton helicity nonflip $\hat H_i$ is closed to $H_i\;\;$:
 $\hat H^i=[H^i-\frac{\xi^2}{1-\xi^2} E^i] + [\tilde H^i,\; (\tilde
 E^i)]$. For the amplitude $M_{\mu' -,\mu +}$ with proton helicity-flip we have:
  $\hat H^i\propto E^i + \xi \tilde
 E^i$

The skewness $\xi$ is related to Bjorken-$x$ by $\xi\simeq x_B/2$.
In the region of small $x \leq 0.01$  gluons give the dominant
contribution. At larger $x \sim 0.2$  the   quark contribution
plays an important role \cite{gk06}.

The hard amplitude is calculated   using the $k$- dependent wave
function \cite{koerner} that contains the leading and higher twist
terms describing the longitudinally and transversally polarized
vector mesons, respectively. The subprocess amplitude is
calculated within the MPA \cite{sterman}. The  amplitude ${\cal
H}^V$ is represented as  the contraction of the hard
  part $F$, which is calculated perturbatively, and the
non-perturbative meson  wave function $ \phi_V$
\begin{equation}\label{hsaml}
  {\cal H}^V_{\mu'+,\mu +}\,=
\,\frac{2\pi \als(\mu_R)}
           {\sqrt{2N_c}} \,\int_0^1 d\tau\,\int \frac{d^{\,2} \vk}{16\pi^3}
            \phi_{V}(\tau,k^2_\perp)\;
                  F_{\mu^\prime\mu}^\pm (\xb,\xi,\tau,Q^2,k^2_\perp).
\end{equation}
In the hard subprocess $F$ we keep the $k^2_\perp$ terms in the
denominators of LL and TT transitions and in the numerator of the
TT amplitude. The gluonic corrections are treated in the form of
the Sudakov factors. The resummation and exponentiation of the
Sudakov corrections can be performed  in the impact parameter
space. We use the Fourier transformation to switch from $\vk$ to
$\vbs$ space. The wave function is chosen in the Gaussian form
\begin{equation}\label{wave-l}
  \phi_V(\vk,\tau)\,=\, 8\pi^2\sqrt{2N_c}\, f_V a^2_V
       \, \exp{\left[-a^2_V\, \frac{\vk^{\,2}}{\tau\bar{\tau}}\right]}\,.
\end{equation}
 Here $\bar{\tau}=1-\tau$, $f_V$ is the
decay coupling constant. The $a_V$ parameter determines the mean
value of the quark transverse momentum $<\vk^{\,2}>$  in the
vector meson. The $a_V$ values are  different for longitudinal and
transverse polarization of the meson \cite{gk05,gk06}.

To estimate  GPDs, we use the double distribution representation
\cite{mus99}
\begin{equation}
  H_i(\xb,\xi,t) =  \int_{-1}
     ^{1}\, d\beta \int_{-1+|\beta|}
     ^{1-|\beta|}\, d\alpha \delta(\beta+ \xi \, \alpha - \xb)
\, f_i(\beta,\alpha,t)
\end{equation}
which connects, through the double distribution function $f$, GPDs
with parton distributions (PDFs)
\begin{equation}\label{ddf}
f_i(\beta,\alpha,t)= e^{b_i\,t}\, |\beta|^{-\alpha't}\,
h_i(\beta)\,
                   \frac{\Gamma(2n_i+2)}{2^{2n_i+1}\,\Gamma^2(n_i+1)}
                   \,\frac{[(1-|\beta|)^2-\alpha^2]^{n_i}}
                           {(1-|\beta|)^{2n_i+1}}.
                          \end{equation}
The  Regge motivated anzats is used here to model $t$ dependences
of PDFs. Here we assume that the Regge trajectories are  linear
functions of $t$  at small momentum transfer
$\alpha_i=\alpha_i(0)+\alpha'_i t$.

The powers $n_i$ (i= gluon, sea, valence contributions) and the
functions $h_i(\beta,t)$ which are expressed in PDFs terms  are
determined by
\begin{eqnarray}\label{pdf}
& h_g(\beta,0)=|\beta|g(|\beta|),& n_g=2;\nonumber\\
& h_{sea}^q(\beta,0)=q_{sea}(|\beta|) \mbox{sign}(\beta), & n_{sea}=2;\nonumber\\
& h_{val}^q(\beta,0)=q_{val}(\beta) \Theta(\beta), & n_{val}=1.
\end{eqnarray}
Here $g$ and $q$ are the ordinary gluon and quark PDF. To
calculate GPDs, we use the CTEQ6 fits of  PDFs for gluon, valence
quarks and sea \cite{CTEQ}. Note that the $u$ and $d$  seas have
very similar $\beta$ and $Q^2$ dependences. This is not the case
for the strange sea. In agrement with CTEQ6 PDFs  we suppose the
flavor asymmetric quark sea $H^u_{sea} = H^d_{sea} = \kappa_s
H^s_{sea}$ \cite{gk07q}, with
\begin{equation}\label{kapp}
\kappa_s=1+0.68/(1+0.52 \ln(Q^2/Q^2_0)).
\end{equation}

\begin{figure}[h!]
\begin{center}
\begin{tabular}{ccc}
{\includegraphics[width=4.5cm,height=4.2cm]{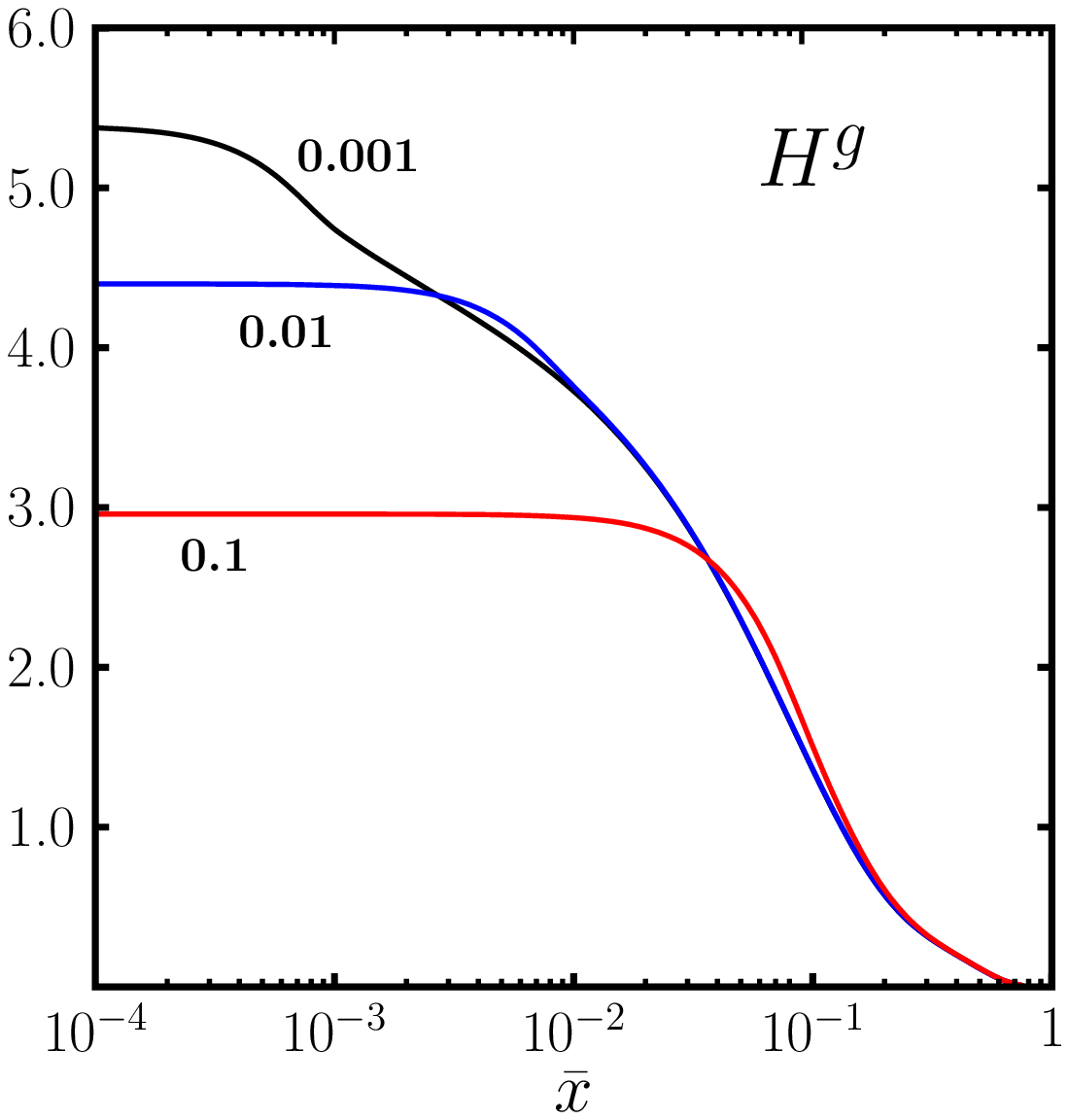}}&
\includegraphics[width=4.5cm,height=4.2cm]{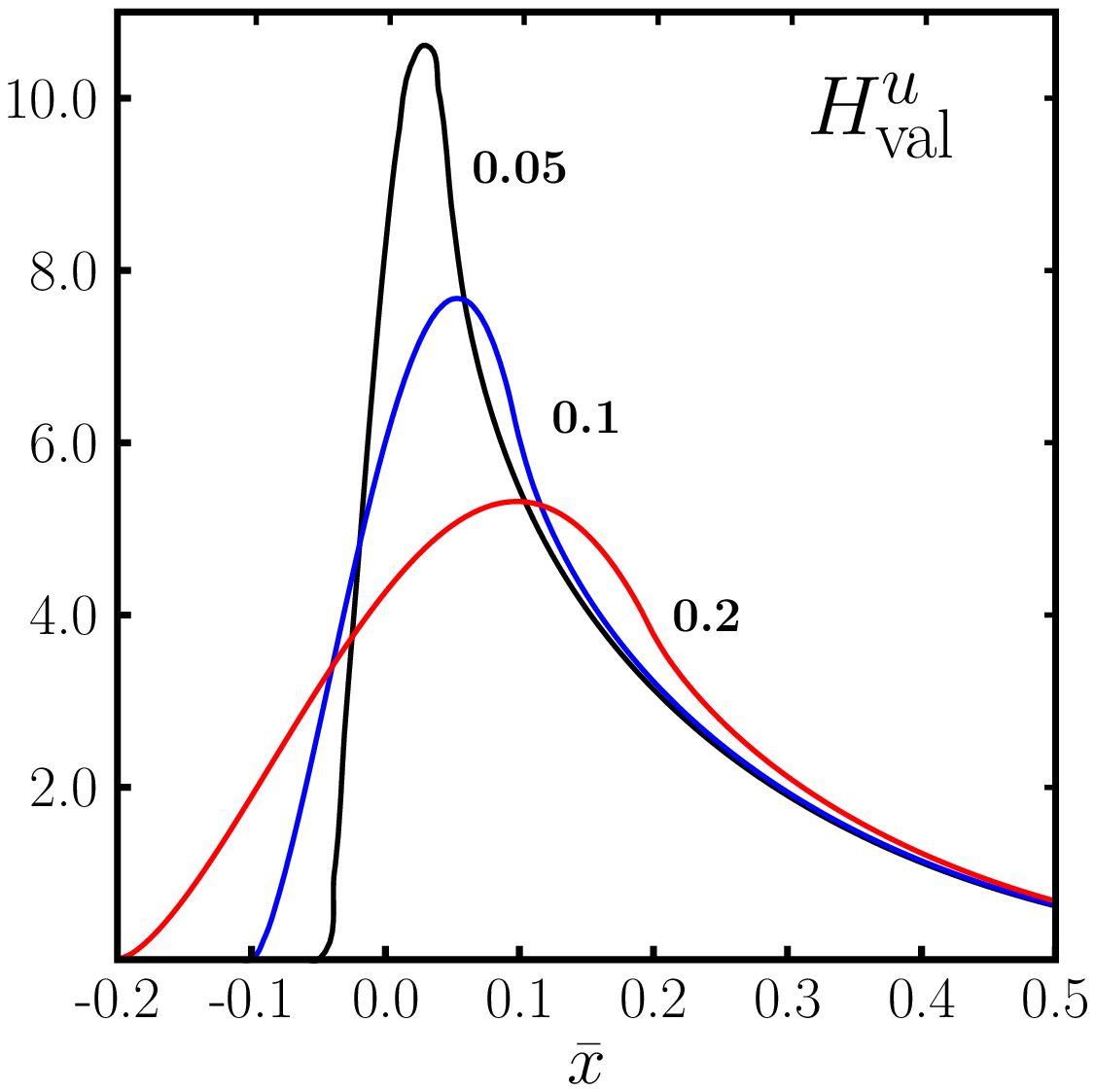}&
\includegraphics[width=4.5cm,height=4.2cm]{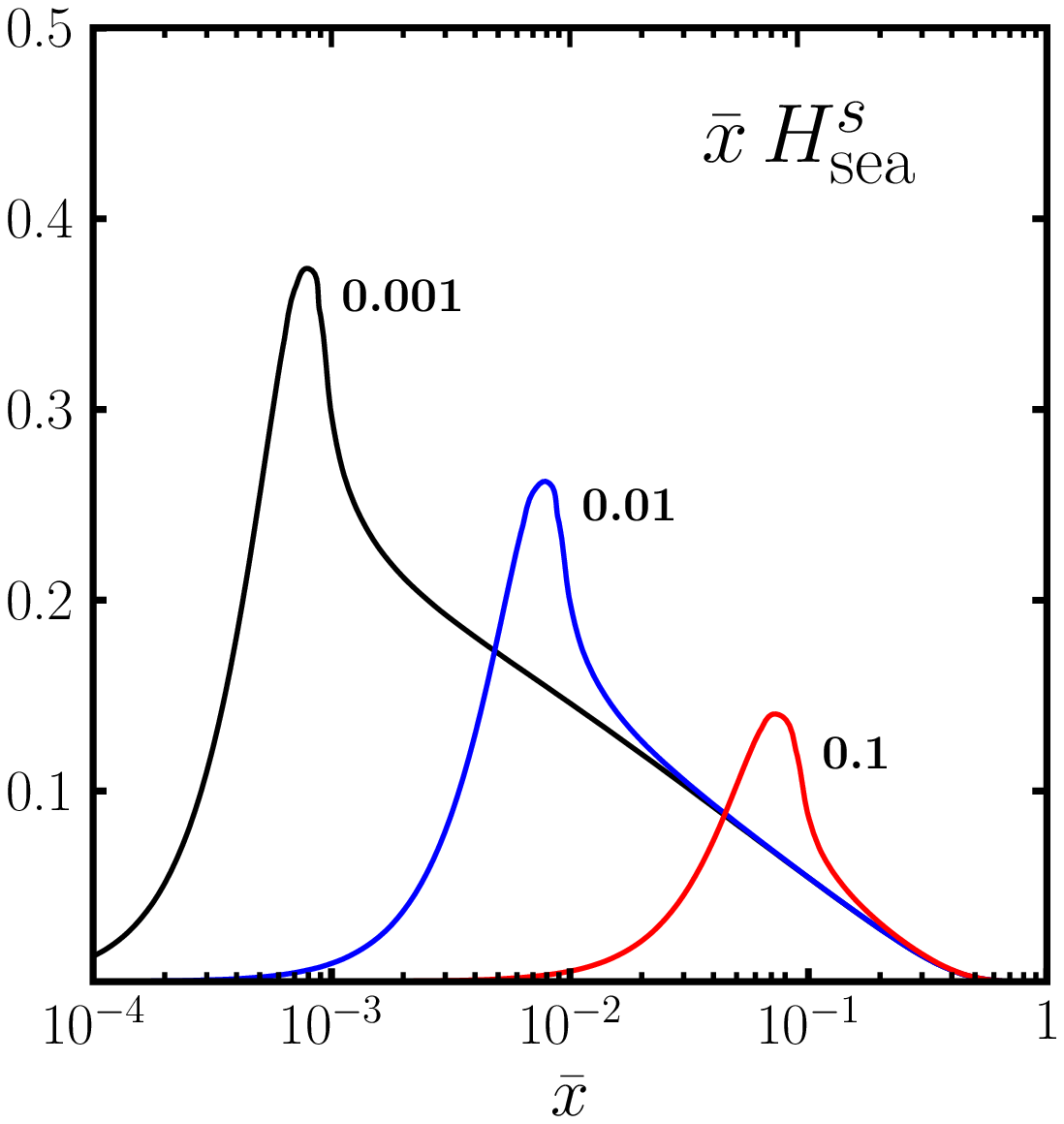}\\
\textbf{(a)} & \textbf{(b)}& \textbf{(c)}
\end{tabular}
\end{center}
\caption{ GPDs {\bf(a)} $H^g$, {\bf(b)} $H^u_{val}$ {\bf(c)}
$H^s_{sea}$ for some values of skewness via $x$. GPDs are shown at
$t =0$ and scale $Q^2=4\mbox{GeV}^2$.}
\end{figure}
The model results for the gluon and valence quark and strange sea
GPDs for the three $\xi$ values are shown in Fig. 1.

\section{Cross section and spin observables in vector meson electroproduction}
In this section, we  analyse vector meson electroproduction and
consider the cross section and spin observables. In the previous
section, the method of amplitudes and GPDs calculation was
described. The  $a_V$ parameters in the wave function were
determined from the best description of the cross section. They
can be found in \cite{gk05,gk06,gk07q} together with other
parameters of the model. We consider the gluon, sea and quark GPDs
contribution to the amplitude. This permits us to analyse vector
meson production from low $x$  to moderate values of $x$ ($ \sim
0.2$) typical of HERMES and COMPASS.

At HERA energies we consider gluon and sea contributions. In this
energy range,  the valence quark effects can be practically
neglected. The cross section for the $\gamma^* p \to \rho p$
production integrated over $t$ is shown in Fig. 2.a. Good
agreement with $Q^2$ dependences of H1 \cite{h1} and ZEUS
\cite{zeus} data is found. The model results for the $\phi$
production cross section shown in Fig. 2.b describe  well the
experimental data \cite{h1,zeus}. The theoretical uncertainties of
our results for the cross sections are estimated from the Hessian
errors of the CTEQ6 PDFs, see the discussion in \ci{gk06,gk07q}.
Due to neglected power corrections of order $m^2/Q^2$ and $-t/Q^2$
and the possibility of large higher order perturbative QCD
corrections \ci{kugler07}, we practically do not provide results
for $Q^2<3\,\gev^2$.

\begin{figure}[h!]
\begin{center}
\begin{tabular}{cc}
\includegraphics[width=6.8cm]{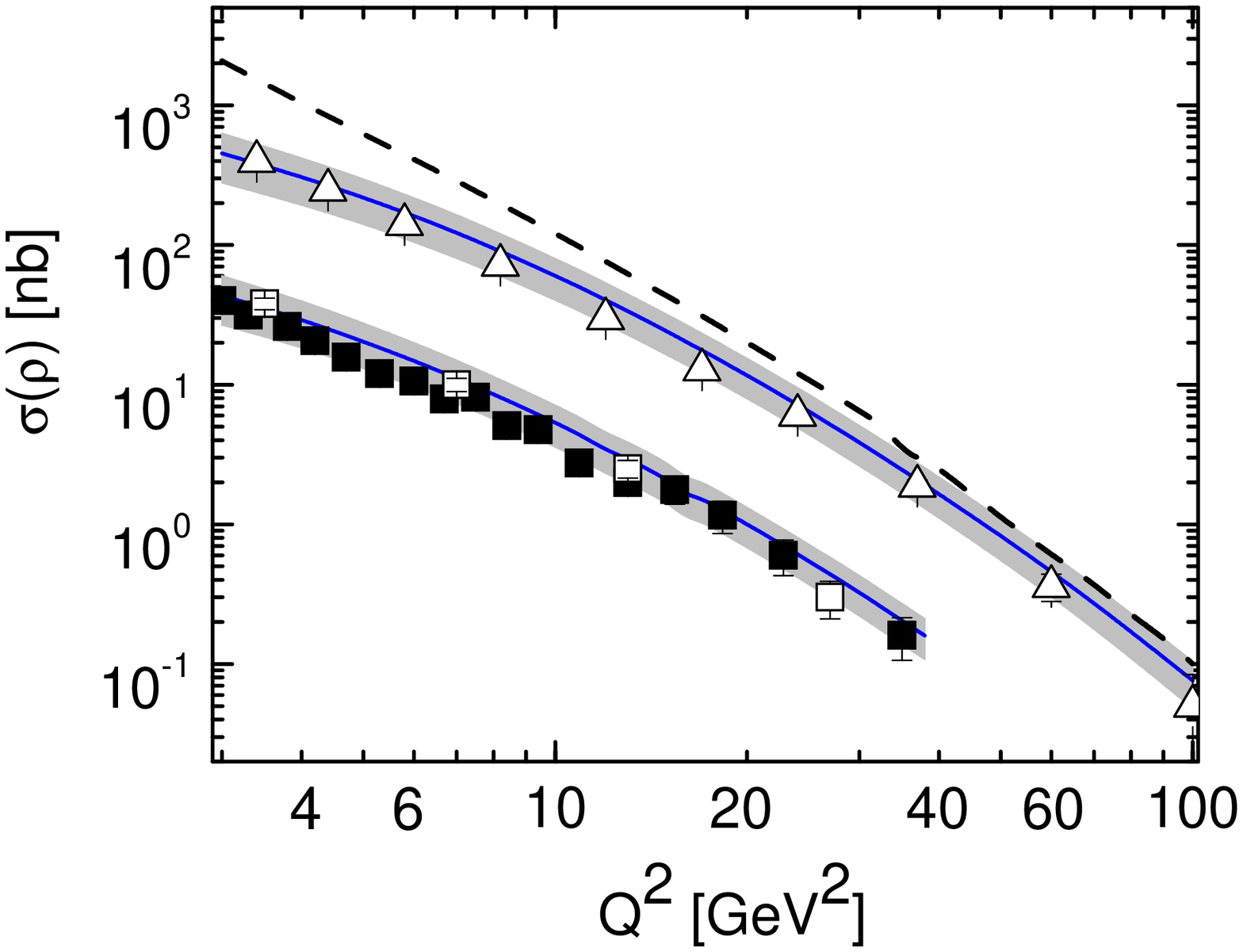}&
\includegraphics[width=6.8cm]{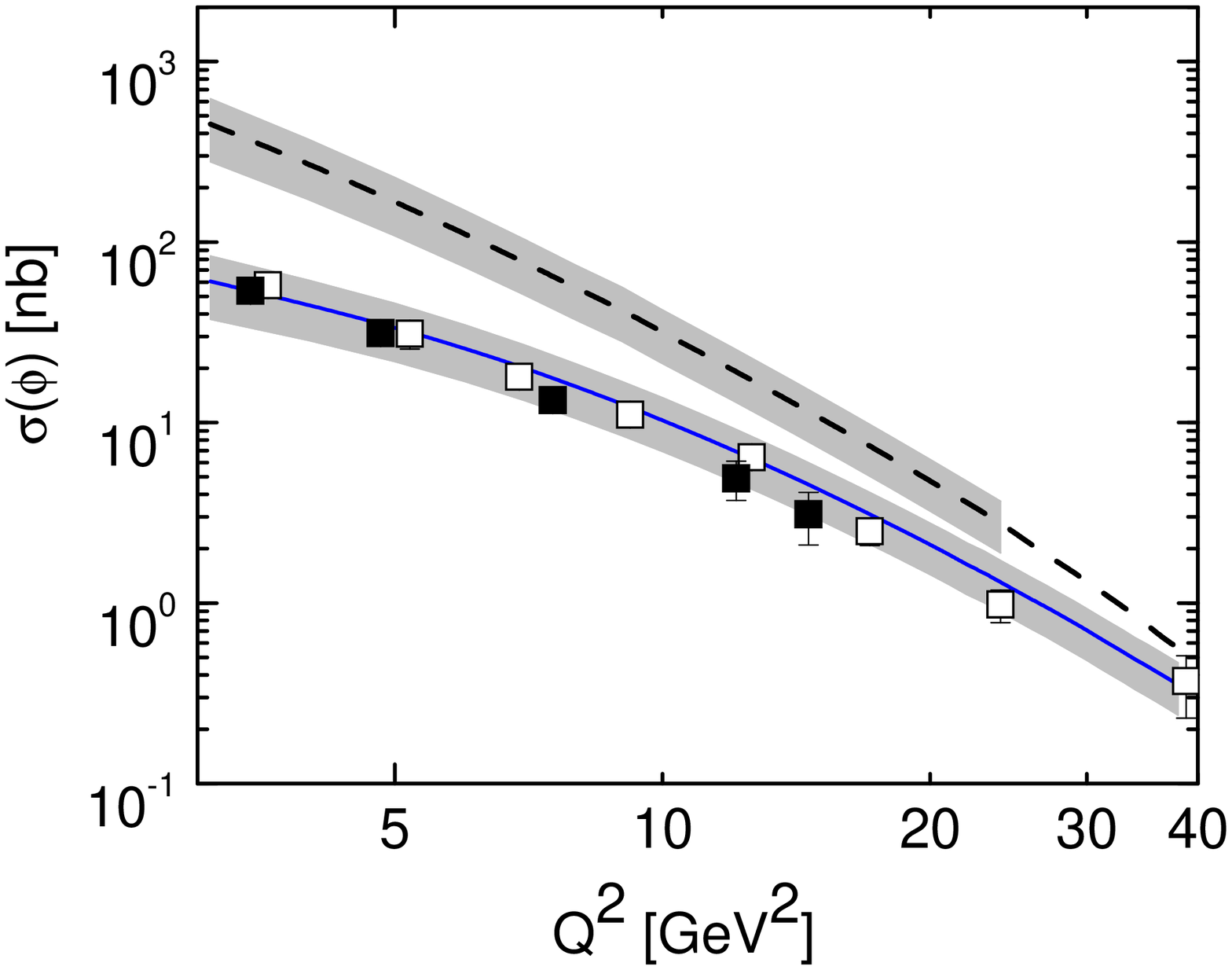}\\
\textbf{(a)} & \textbf{(b)}
\end{tabular}
\end{center}
\caption{  {\bf(a)} Longitudinal cross sections of $\rho$
production at $W=90 \mbox{GeV}$-open triangles and  $W=75
\mbox{GeV}$ scaled by the factor 10 with error band from CTEQ6
PDFs uncertainties. {\bf(b)} Longitudinal cross sections of $\phi$
production at $W=75 \mbox{GeV}$. Data are from H1  -solid symbols
and ZEUS -open symbols. Dashed line - leading twist result. }
\end{figure}

\begin{figure}[h!]
\begin{center}
\begin{tabular}{cc}
\includegraphics[width=6.8cm,height=4.9cm]{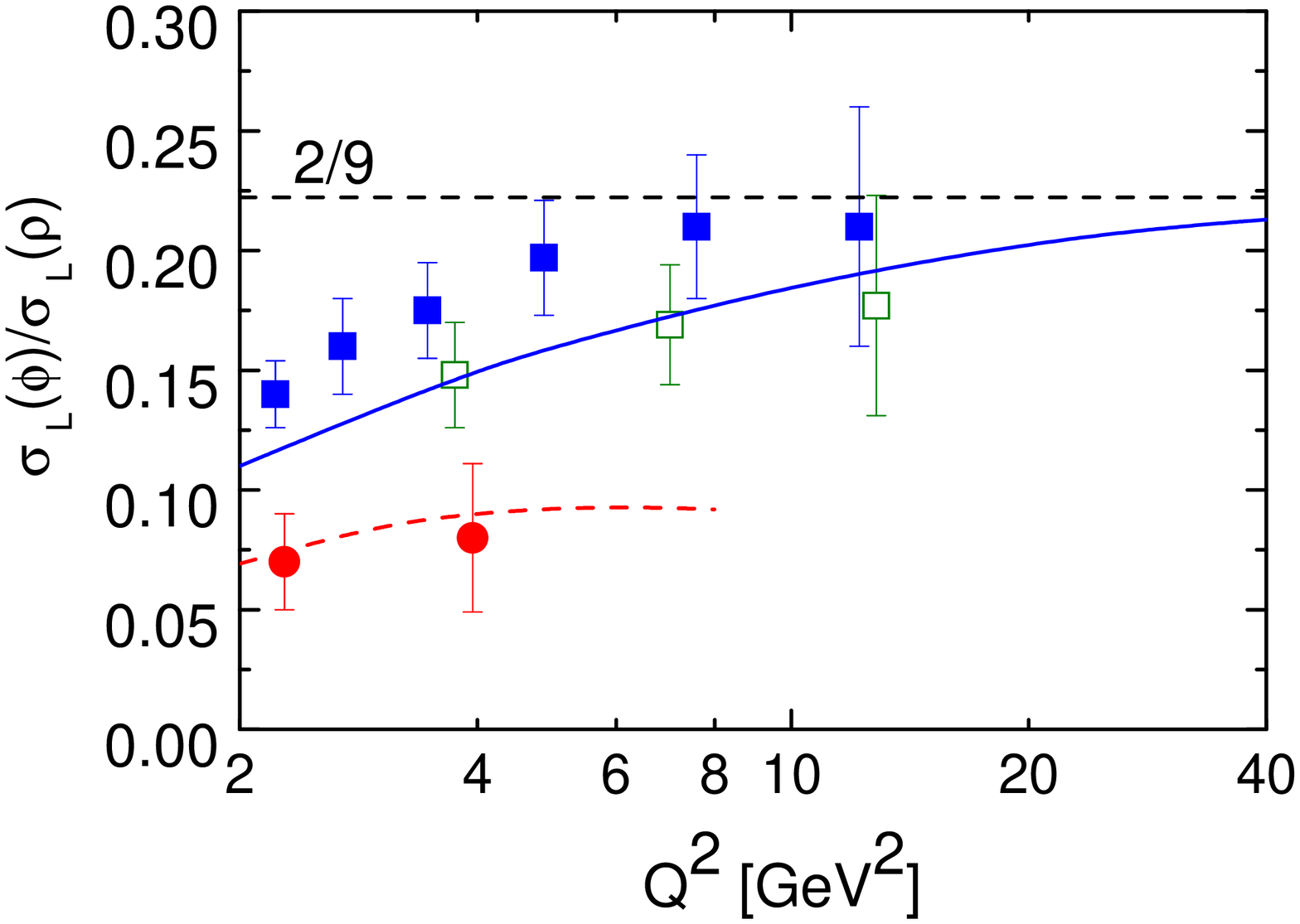}&
\includegraphics[width=6.8cm,height=4.9cm]{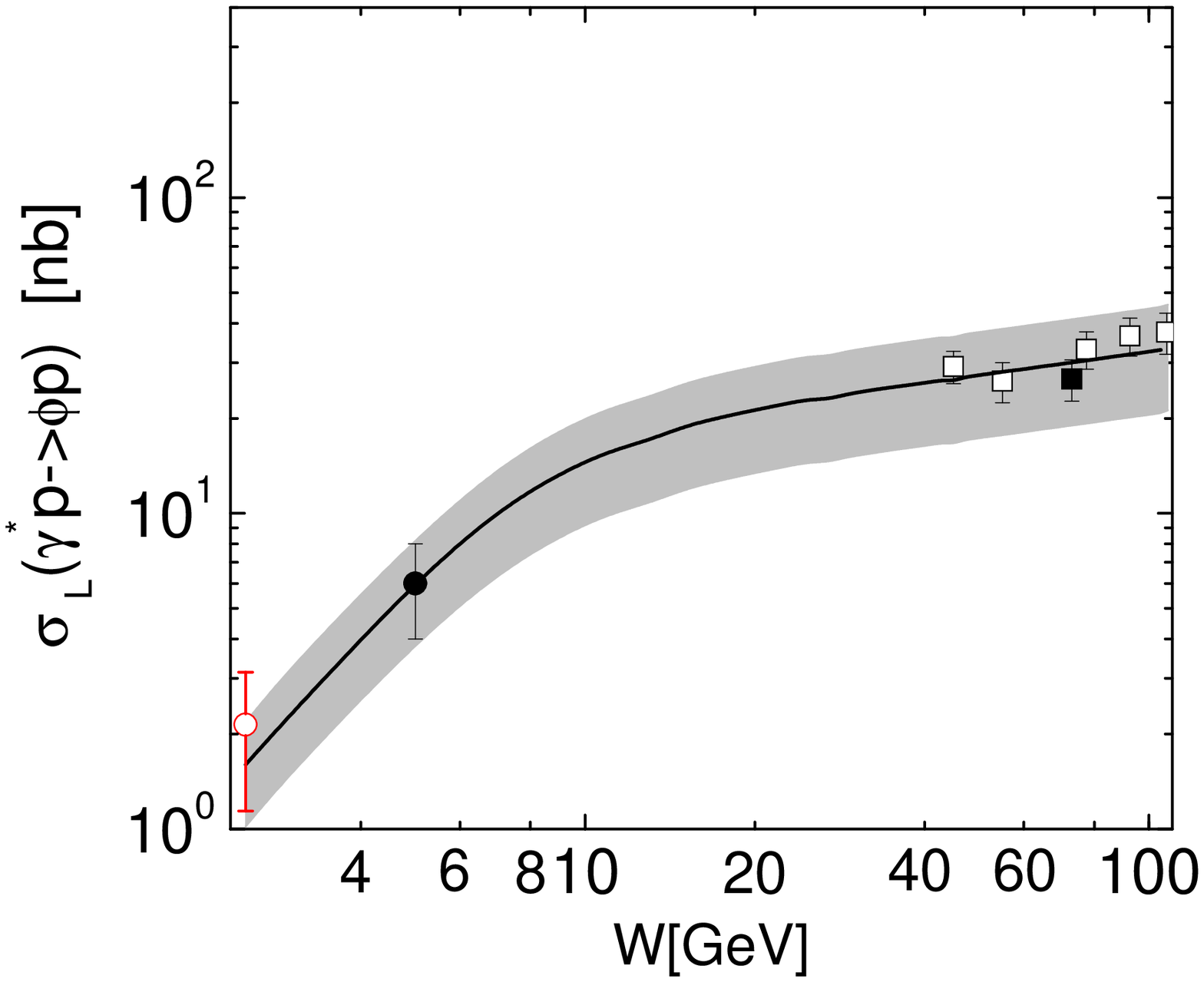}\\
{\bf(a)}& {\bf(b)}
\end{tabular}
\end{center}
\caption{ {\bf(a)} The  ratio of cross sections
$\sigma_\phi/\sigma_\rho$ at
    HERA energies- full line and HERMES- dashed line. Data are from H1  -solid, ZEUS
 -open squares, HERMES solid circles. {\bf(b)}The longitudinal cross section for
 $\phi$ at
$Q^2=3.8\,\mbox{GeV}^2$. Data: HERMES, ZEUS, H1, open circle- CLAS
data point.}
\end{figure}

The leading twist results, which do not take into account effects
of a transverse quark motion, are presented in Fig. 2, too. One
can see that the $k_\perp^2/Q^2$ corrections in the hard amplitude
 are extremely important at low $Q^2$. They decrease
the cross section by a factor of about 10 at $Q^2 \sim
3\mbox{GeV}^2$. We see that $k_\perp^2/Q^2$ corrections are
considerable in the cross section for $Q^2<20 \mbox{GeV}^2$.

It can be shown that for the  flavor symmetric sea the ratio of
the $\sigma_\phi/\sigma_\rho$ cross section should be equal to
2/9. In Fig~3.a, we show the strong violation of  this ratio from
2/9 value  at HERA energies and low $Q^2$ which is caused by the
flavor symmetry breaking (\ref{kapp}) between $\bar u$ and $\bar
s$ sea. The valence quark contribution to $\sigma_\rho$ decreases
this ratio at HERMES energies.

The model results reproduce well the energy dependence of the
$\rho$ and $\phi$ production cross section from HERMES to HERA
energies. It was found that the valence quarks substantially
contribute only at HERMES energies. At lower energies this
contribution becomes small and the cross section decreases with
energy. This is in contradiction with CLAS \cite{clas} results
which show essential increasing of $\sigma_\rho$ for $W<5
\mbox{GeV}$. On the other hand, we found good description of
$\phi$ production at CLAS
 \cite{clas} Fig~3.b. This means that we have problem only with the
valence quark contribution at low CLAS  energies.

The TT amplitude, which is essential for spin observables, is
calculated from HERMES to HERA energies with including  of the
quark effects \cite{gk07q}. We compare our results for spin
observables with experimental data on the ratio of the cross
section with longitudinal and transverse photon polarization
\begin{equation}\label{R}
R=\frac{\sigma_L}{\sigma_T}.
\end{equation}
Our results for the $\rho$  production are shown in Fig. 4.a. We
describe properly available data from H1 and ZEUS experiments
\cite{h1,zeus}. Our results describe well the COMPASS
\cite{compass} and HERMES \cite{hermes} data on the $R$ ratio and
lie lower with respect to HERA data. This shows that the quark
contribution to the transverse and longitudinal amplitude is
different and causes the energy dependence of the $R$-ratio.

\begin{figure}[h!]
\begin{center}
\begin{tabular}{cc}
\includegraphics[width=6.7cm]{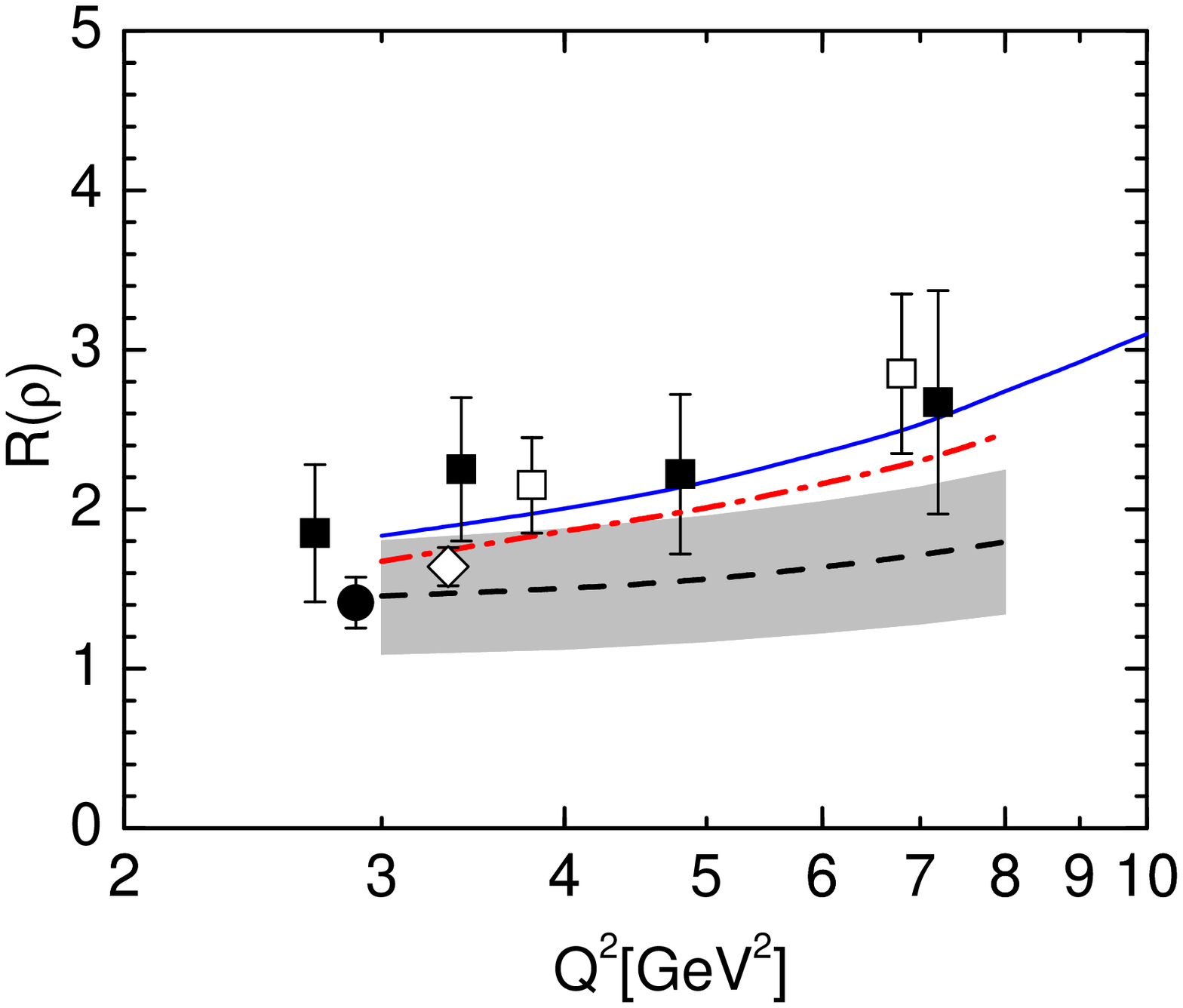}&
\includegraphics[width=6.7cm]{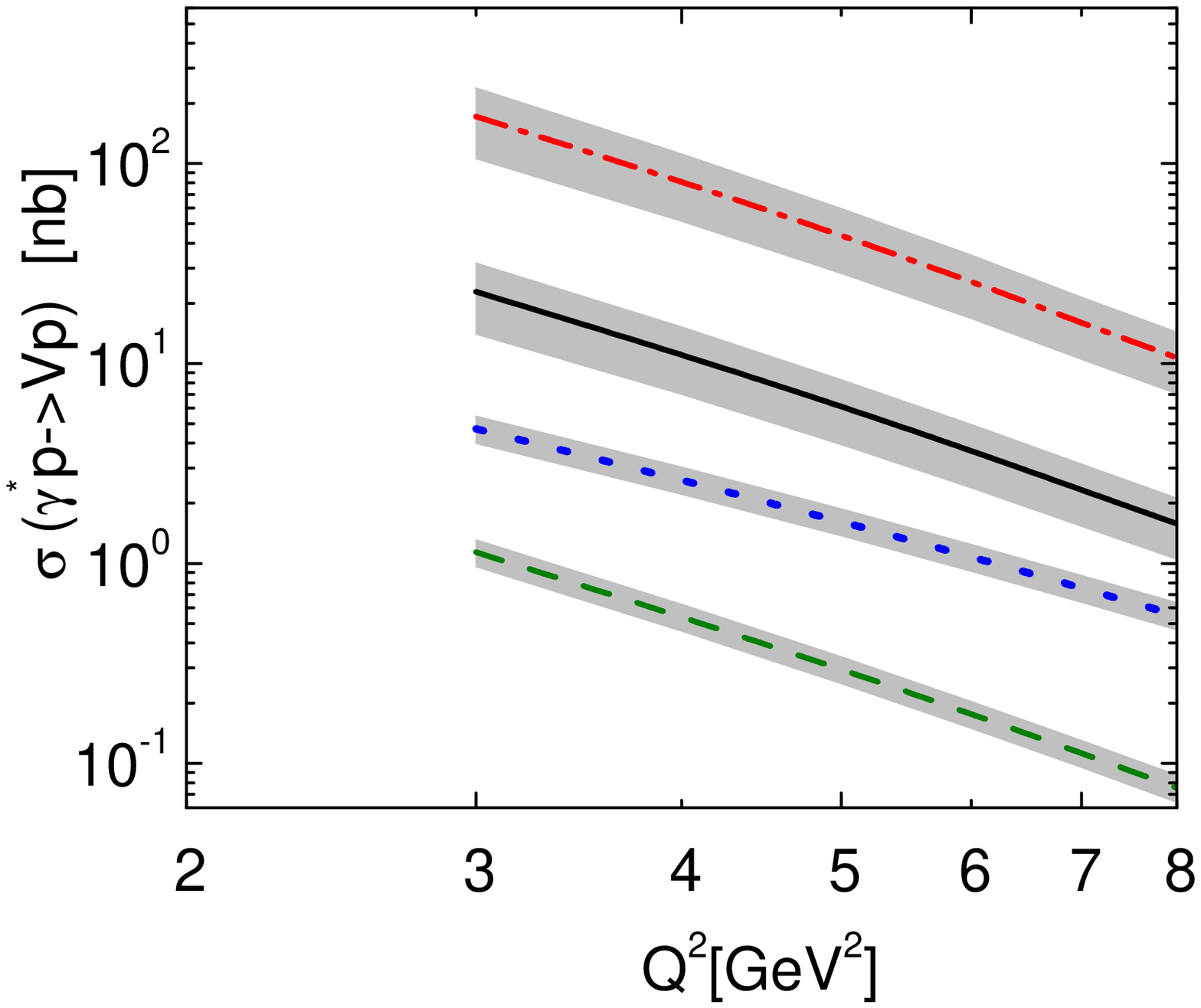}\\
{\bf(a)}& {\bf(b)}
\end{tabular}
\end{center}
\caption{ {\bf(a)} The ratio of longitudinal and transverse cross
sections for $\rho$  production at low  $Q^2$. Full line- HERA ,
 dashed-dotted -COMPASS and dashed- HERMES. {\bf(b)} Predicted cross section
  at HERMES for various mesons.
    Dotted-dashed line $\rho^0$;
    full line $\omega$; dotted line $\rho^+$ and dashed line $K^{* 0}$}
\end{figure}

We have analysed the  spin density matrix elements $r^{04}_{00},
r^1_{1-1}, r^2_{1-1}$ which are sensitive to the LL and TT
amplitudes  and two  SDME  $\mbox{Re} \,r^5_{10}, \mbox{Im}
\,r^6_{10}$ which are expressed in terms of the LL and TT
amplitudes interference. Description of experimental data is
reasonable. The last two SDME are relevant to the phase shift
between the LL and TT amplitudes $\delta_{LT}$. The experimental
analyses show a quite large phase difference $\delta_{LT} \sim
20-30^o$ \cite{hermes}. It is important to note that a similar
phase shift $\delta_{LT} \sim 20^o$ is observed at H1 experiment
\cite{h109}. This shows that such a phase shift should appear in
the the gluon contribution to the LL and TT amplitudes and is not
understood in our model that gives small $\delta_{LT} \sim 2-3^o$
\cite{gk05}.

For different meson production at low energies we can get
information about valence and sea quarks. Really, the quarks
contribute to meson production reactions in different
combinations. For uncharged meson production we have standard GPDs
and find
\begin{equation}\label{quarks}
\rho:\;\; \propto \frac{2}{3} H^u +\frac{1}{3} H^d;\;\;\omega:\;\;
\propto \frac{2}{3} H^u -\frac{1}{3} H^d;\;\;\pi^0:\;\; \propto
\frac{2}{3} \tilde H^u +\frac{1}{3} \tilde H^d.
\end{equation}

For production of charged and strange mesons we have transition $
p\to n$ and $p\to \Sigma$ GPDs:
\begin{equation}\label{quarkspl}
\rho^+:\;\; \propto H^u -H^d;\;\;\pi^+:\;\;\propto \tilde H^u -
\tilde H^d;\;\;K^{*0}:\;\;  \propto  H^d -H^s.
\end{equation}
The same is valid for $E, \tilde E$ contributions. Thus, we can
test these GPDs  in the mentioned reactions.

Predictions for the unseparated integrated over $t$ cross sections
for the $\omega$, $\rho^+$ and $K^{*0}$ electroproduction are
shown in Fig. 4.b where, for comparison,  results for the $\rho^0$
production are also displayed. The cross sections for the
$\rho^0$, $\phi$ and $\omega$ production increase with energy at
fixed $Q^2$ due to considerable contributions from the gluonic
subprocess $\gamma^* g\to Vg$ which grows $\propto
W^{4\delta_g(Q^2)}$ for $\xi\to 0$. This behaviour is to be
contrasted with the cross sections for $\gamma^* p\to
K^{*0}\Sigma^+$ and $\gamma^*p\to\rho^+p$ to which the gluons do
not contribute. These cross sections are predicted to decrease
with energy since the dominant valence quark contributions lead to
$\sigma\propto W^{4(\alpha_{\rm val}(0)-1)}$ as $\xi\to 0$. This
decrease is milder for $K^{*0}$ than for the $\rho^+$ channel due
to the strange quark contribution which has the same energy
dependence as the gluon one. Lack of data for these processes
prevents the verification of our results.

The $A_{UT}$ asymmetry is sensitive to  interferences of the
amplitudes determined by the $E$ and $H$ GPDs.
\begin{equation}\label{aut}
A_{UT}=-2 \frac{\mbox{Im}[M_{+-,++}^*\, M_{++,++}+\varepsilon
M_{0-,0+}^*M_{0+,0+}]}{\sum_{\nu}' [|M_{+ {\nu}',++}|^2
+\varepsilon |M_{0 {\nu}',0+}|^2] }\propto \frac{\mbox{Im}<E>^*\,
<H> }{|<H>|^2}.
\end{equation}
 We  constructed the GPD
$E$ from double distributions and constrained it by the Pauli form
factors of the nucleon, positivity bounds and sum rules. The GPDs
$H$ were taken from our  analysis of the vector meson
electroproduction cross section. Our results for the
$\sin(\phi-\phi_s)$ moment of the $A_{UT}$ asymmetry for the
$\rho^0$ production \cite{gk08} at $W=5 \mbox{GeV}$ are shown in
Fig. 5.a. It can be seen that  results on $t$- dependence of
asymmetry are in agreement with HERMES data \cite{hermesaut}. Our
prediction for  $A_{UT}$ asymmetry of the $\rho^0$ production at
COMPASS is at the level of 1-2\% and reproduces well the
preliminary experimental data \cite{sandacz}.

 Predictions for the $A_{UT}$ asymmetry at $W=10
\mbox{GeV}$ are given for the $\omega$, $\phi$, $\rho^+$, $K^{*0}$
mesons \cite{gk08} in Fig~5.b. It can be seen that we predicted
not small negative asymmetry for the $\omega$ and large positive
asymmetry for the $\rho^+$ production. In these reactions, the
valence $u$ and $d$ quark GPDs contribute to the production
amplitude in combination $\sim E^u-E^d$ (without the corresponding
coefficients- see (\ref{quarks})) and do not compensate each other
because GPDs $E^u$ and $E^d$ have different signs. The opposite
case is for the $\rho^0$ production where one has the $\sim 2/3
E^u+1/3 E^d$ contribution to the amplitude and valence quarks
compensate each other essentially. As a result, the $A_{UT}$
asymmetry for $\rho^0$ is predicted to be quite small.
Unfortunately, it is difficult to analyse  the $A_{UT}$ asymmetry
for $\omega$ and $\rho^+$ production experimentally because the
cross section for  these reactions is much smaller compared to
$\rho^0$, Fig.~4.b.

\begin{figure}[h!]
\begin{center}
\begin{tabular}{ccc}
\includegraphics[width=6.8cm,height=5.3cm]{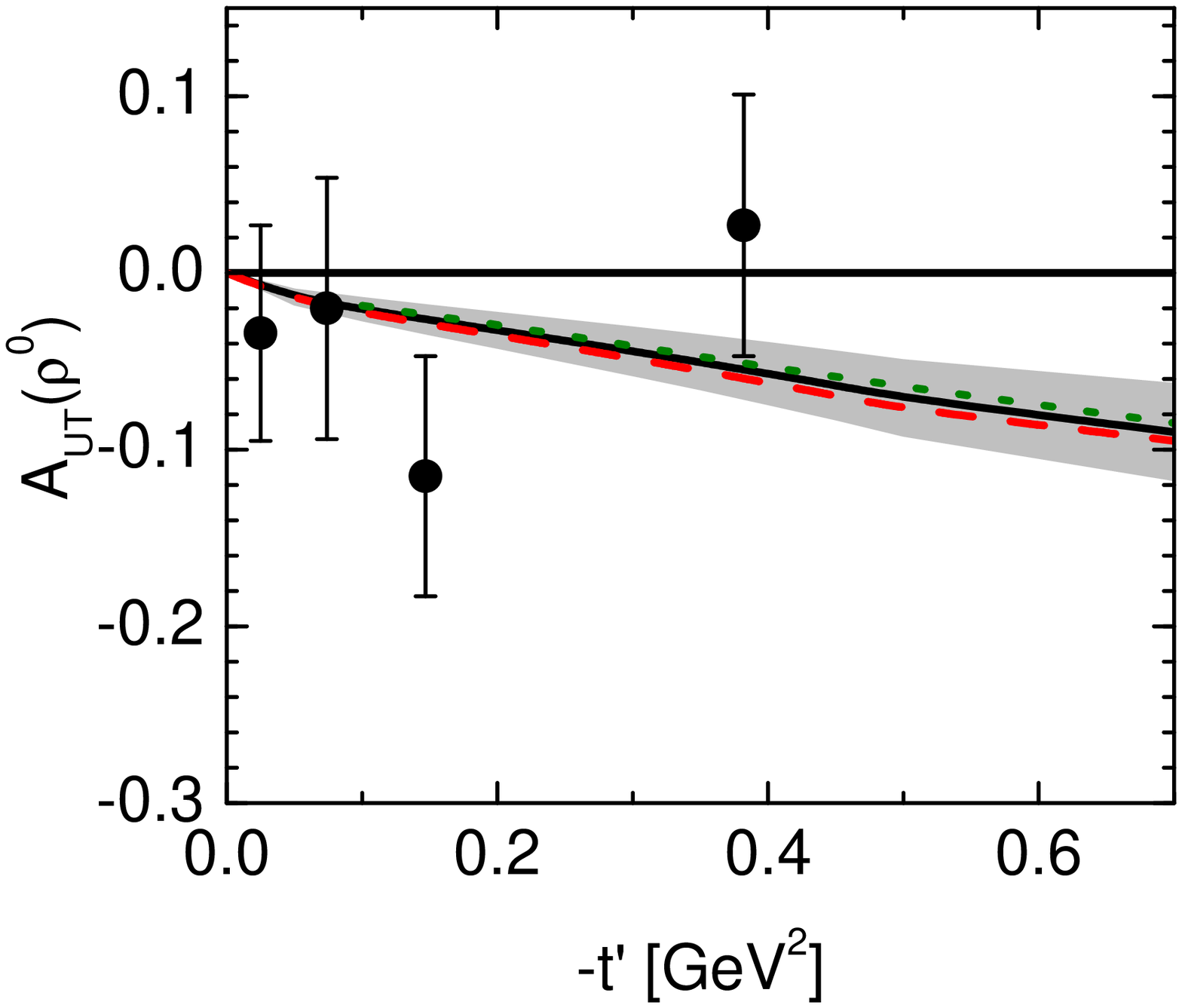}&
\includegraphics[width=6.8cm,height=5.3cm]{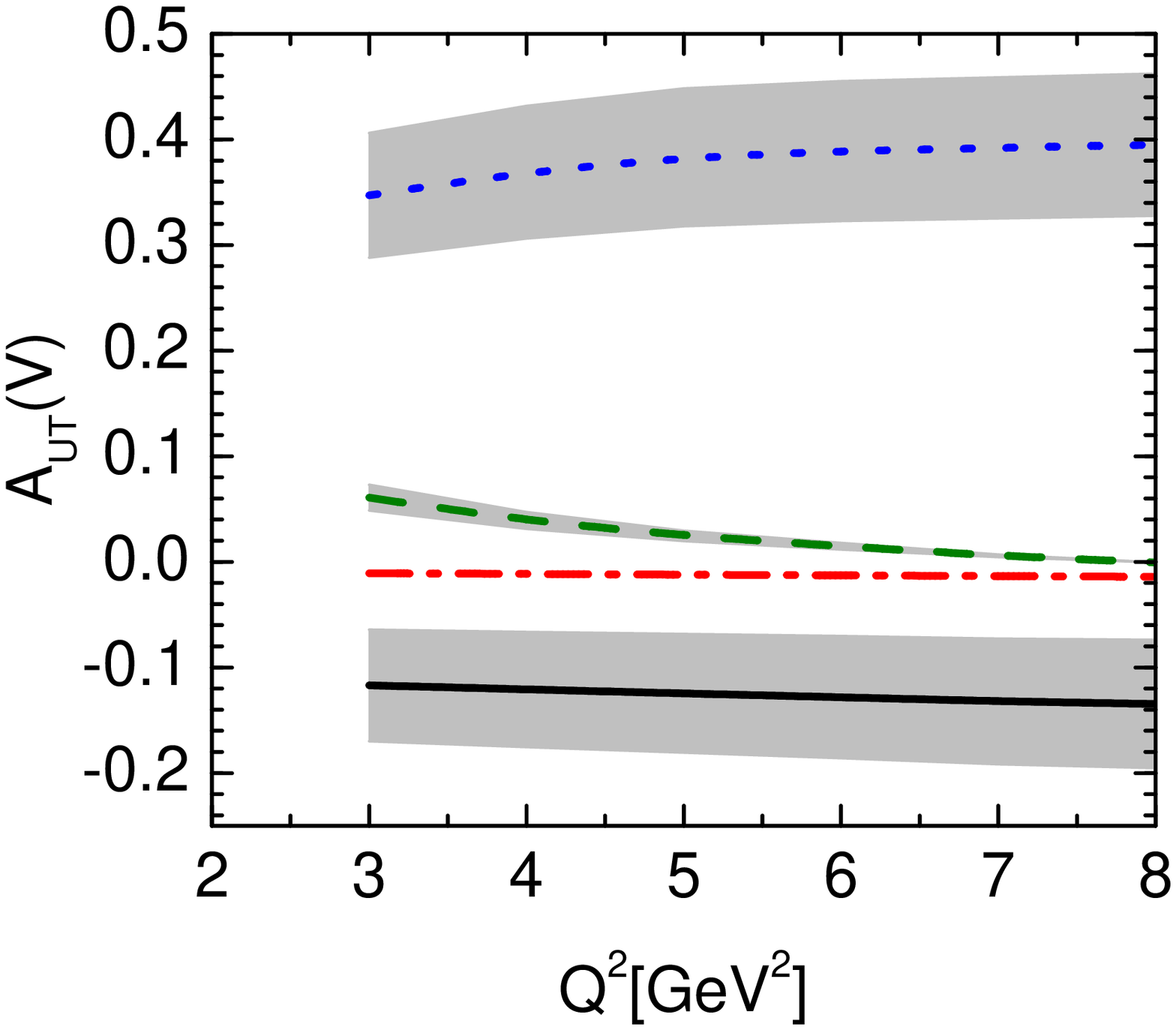}\\
{\bf(a)}& {\bf(b)}
\end{tabular}
\end{center}
\caption{ {\bf(a)}  Predictions for $A_{UT}$ asymmetry  $W=5
\mbox{GeV}$ and $Q^2=2 \mbox{GeV}^2$. Data are from HERMES.
{\bf(b)} Predicted $A_{UT}$ asymmetry at COMPASS for various
mesons. Lines are the same as in Fig. 4b}
\end{figure}
\section{Electroproduction of pions}
Exclusive electroproduction of $\pi^+$ mesons at large $Q^2$
  was investigated within the handbag approach \cite{gk09}. It was
found that the process cannot be understood without taking into
account pion exchange with the full experimentally measured
electromagnetic form factor of the pion.  In addition to the pion
pole and the GPDs $\widetilde{H}$ and $\widetilde{E}$ a twist-3
contribution to the amplitude ${\cal M}_{0-,++}$ is required to
describe the polarization data. In order to estimate this effect,
we use a mechanism that consists of the helicity-flip GPD $H_T$
and the twist-3 pion wave function. For the latter  only the
two-particle partonic states are considered. The subprocess
amplitudes are calculated within the modified perturbative
approach. Higher order perturbative corrections other than those
included in the Sudakov factor  are not taken into account
\cite{gk09}.

In Fig. 6.a, we show the partial cross sections. The longitudinal
one is large at small $t'$ and decreases rapidly with growing
$-t'$. This is caused by the large pion pole contribution to the
longitudinal amplitude. The transverse cross section is quite
large because of the twist-3 contribution.

 We show our results  in Fig. 6.b for the $\sin{(\phi-\phi_s)}$ moment
of the $A_{UT}$ asymmetry in $\pi^+$ production. Large negative
value of asymmetry at small $-t'$ is caused by the pion pole
contribution. The change of sign in asymmetry is produced by
compensation of the pion pole and  $\tilde E$ contributions
\cite{gk09}. Description of HERMES \cite{hermespi} experimental
data is quite good.
\begin{figure}[h!]
\begin{center}
\begin{tabular}{cc}
\includegraphics[width=6.8cm,height=5.3cm]{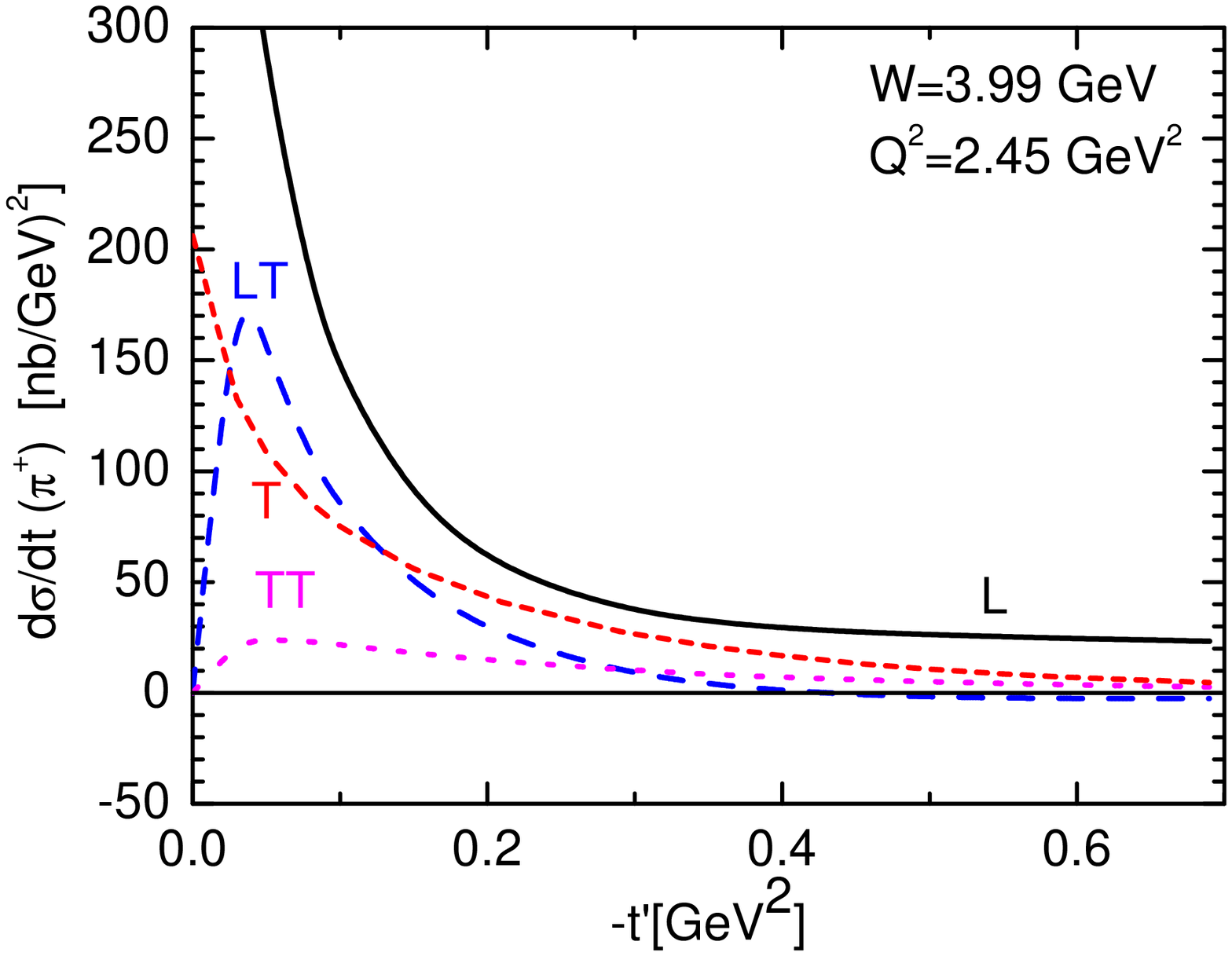}&
\includegraphics[width=6.8cm,height=5.3cm]{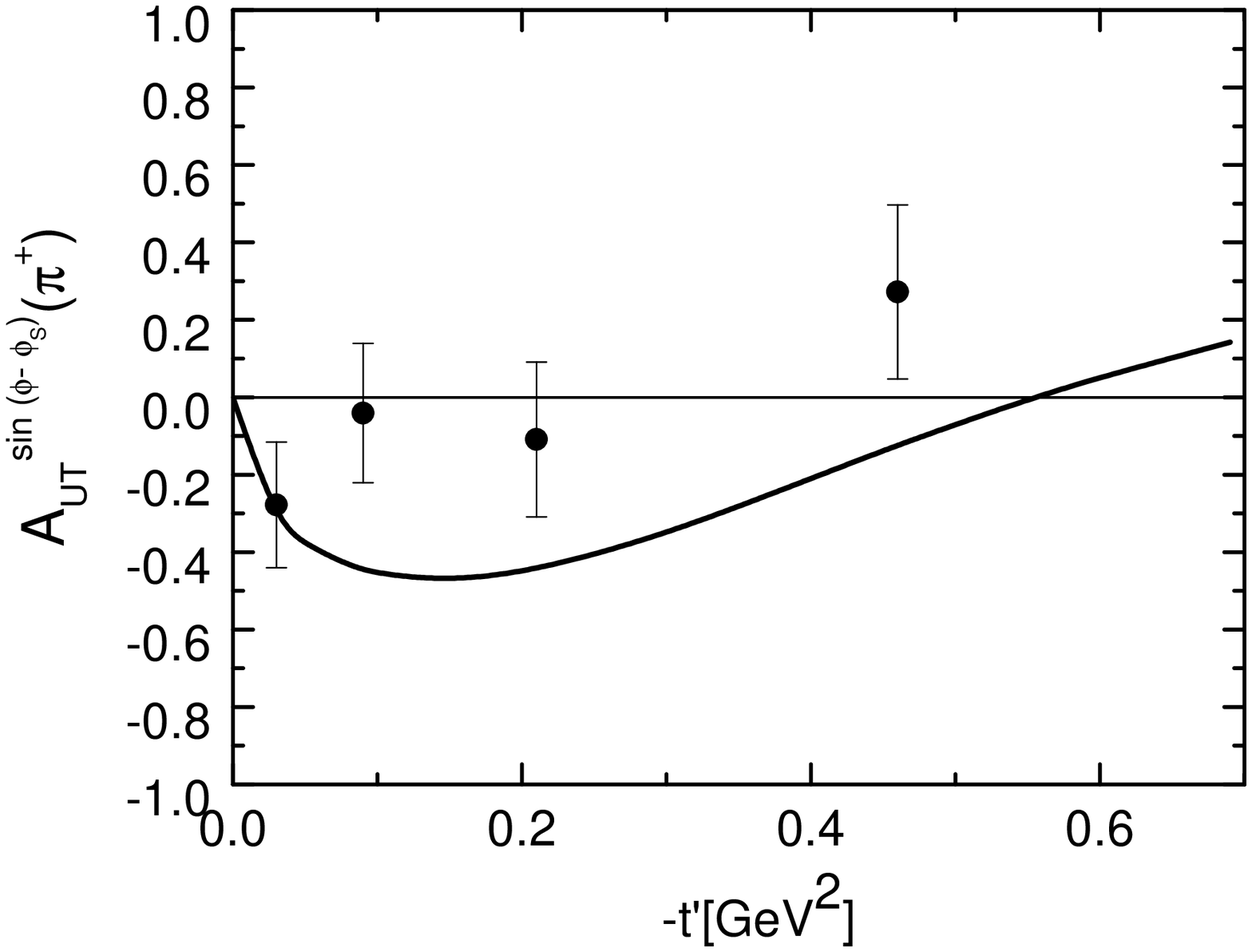}\\
{\bf(a)}& {\bf(b)}
\end{tabular}
\end{center}
\caption{ {\bf(a)} The partial cross section $d \sigma_L/dt, d
\sigma_T/dt, d \sigma_{LT}/dt, d \sigma_{TT}/dt$. {\bf(b)}$A_{UT}$
asymmetry of $\pi^+$ production at $W=4.1 \mbox{GeV}$ and $Q^2=2.4
\mbox{GeV}^2$ with HERMES data.}
\end{figure}
In Fig. 7, our predictions for the $\pi^0$ production are shown.
The cross section is mach smaller with respect to the $\pi^+$ case
(Fig. 7a). This is understandable because the pion pole
contribution is absent in the $\pi^0$ production. The longitudinal
cross section gives the predominant contribution. Essential
moments of $A_{UT} $ asymmetry are shown in Fig. 7.b. The
$\sin(\phi-\phi_s)$ moment of asymmetry is positive. This is
caused by the absence of pion pole contribution in the $\pi^0$
production. The values of asymmetry are large because the
corresponding cross section is small.
\begin{figure}[h!]
\begin{center}
\begin{tabular}{cc}
\includegraphics[width=6.8cm,height=5.3cm]{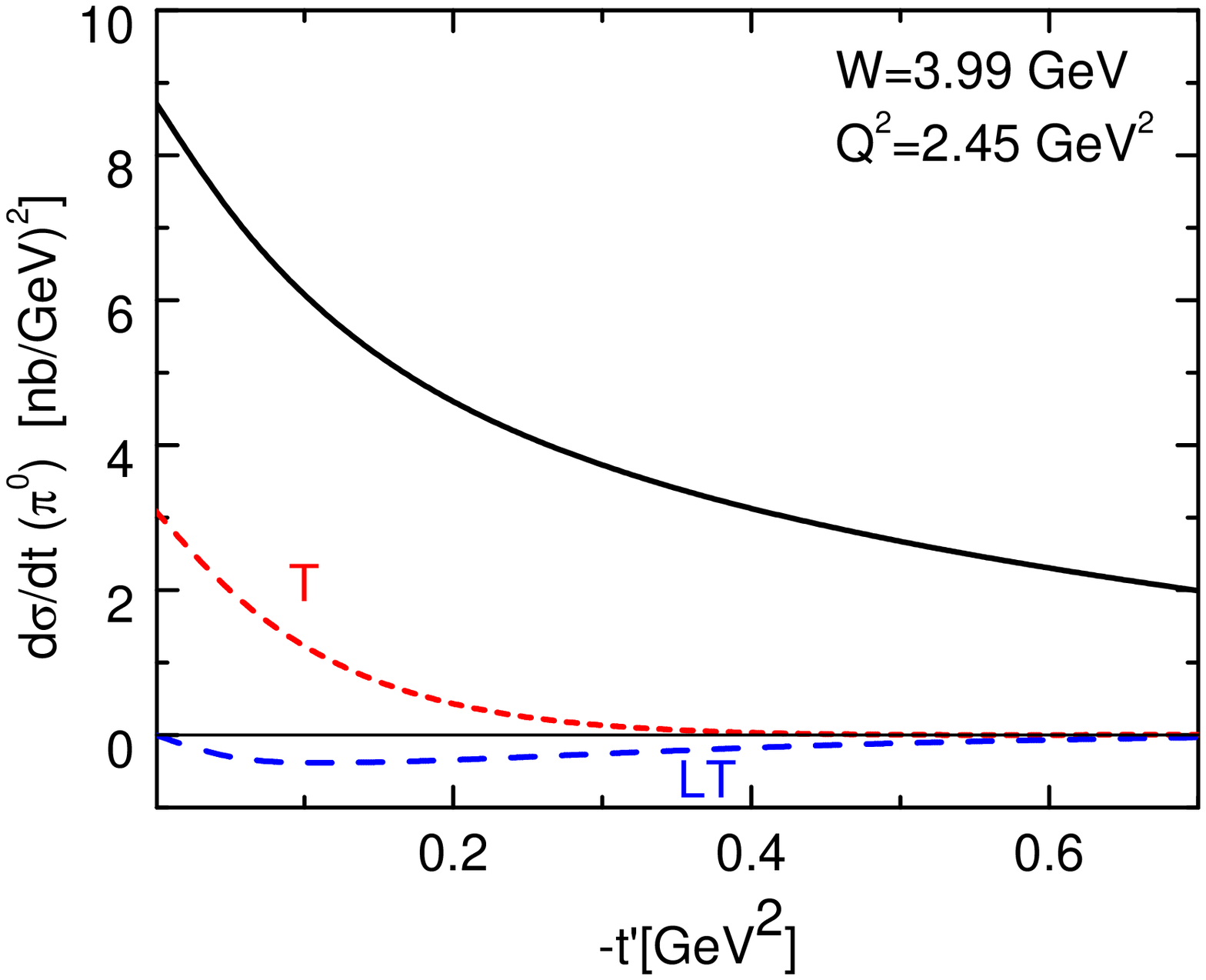}&
\includegraphics[width=6.8cm,height=5.3cm]{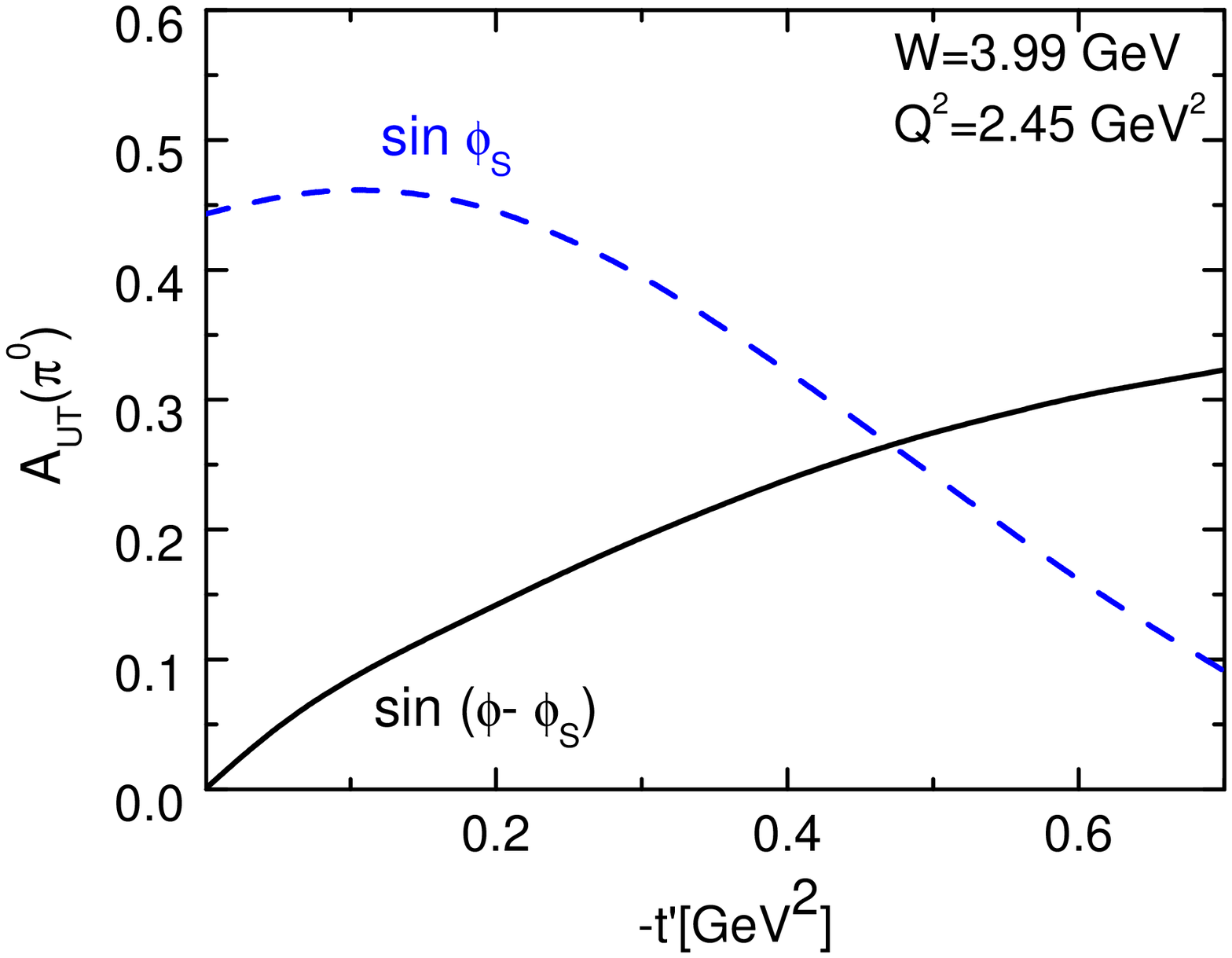}\\
{\bf(a)}& {\bf(b)}
\end{tabular}
\end{center}
\caption{ {\bf(a)} Cross section of the $\pi^0$ production.
Unseparated cross section- full line, transverse-dotted, and
longitudinal-transverse  cross section-dashed line. {\bf(b)}
Predictions for the $\sin{(\phi_s)}$ and $\sin{(\phi-\phi_s)}$
moments of the $A_{UT}$ asymmetry for the $\pi^0$ production at
HERMES }
\end{figure}
 \section{Conclusion and Summary}
In this report, we have studied the light  meson electroproduction
 within the handbag approach.  The amplitude at large $Q^2$
factorizes in the model into a hard subprocess  and GPDs. The MPA,
where the transverse quark motion and the Sudakov factors were
taken into account, was used to calculate the hard subprocess
amplitude. The transverse quark momenta regularize the end-point
singularities in the amplitudes with transversally polarized
photons. This gives a possibility to calculate the TT amplitude
and study spin effects in the vector meson production in our
model. The $k_\perp^2/Q^2$ corrections in the propagators decrease
the cross section by a factor of about 10 at $Q^2 \sim
3\mbox{GeV}^2$. As a result, we describe  the cross section at
quite low $Q^2$.

In the model, a good description of the cross section from HERMES
to HERA energies \cite{gk06} is observed. The gluon and sea
contributions are predominated at energies $W \geq 10 \mbox{GeV}$,
while the valence quarks are essential only at HERMES energies. At
lower energies  the cross section of the $\rho$ and $\phi$
production decreases. This is in agrement with CLAS results for
the $\phi$ production but in contradiction with  essential
increasing of $\sigma_\rho$ for $W<5 \mbox{GeV}$. Thus, the model
describes correctly the gluon and sea contributions up to low CLAS
energies and has some problems with valence quarks.

The model provides a good description of the spin effects,
including the $R$ ratio and SDME for the light meson production in
a wide energy range \cite{gk05,gk06,gk07q}.
 We would like to point out that study of SDME gives  important
information on different  $\gamma \to V$ hard amplitudes. The data
show a quite large phase difference $\delta_{LT} \sim 20-30^o$
\cite{hermes}. Our model gives quite small  $\delta_{LT}$. To
solve this problem, further theoretical and experimental
investigations are necessary.

In the model, the predictions for the cross sections and $A_{UT}$
asymmetries for the $\rho^0$, $\omega$, $\rho^+$ and $K^{*0}$
electroproduction \cite{gk08} were found which give possibility to
test the $H$ and $E$ GPDs for valence and sea quarks. The
experimental data which are available only for the $\rho^0$
production are described well.

The first results on pion production in the model \cite{gk09} were
presented. This reaction give access to $\tilde H$ and $\tilde E$
GPDs. Further theoretical and experimental study of these
precesses is needed. Thus, we can conclude that the gluon and
various quark GPDs can be probed in the meson electroproduction.

 \bigskip

 This work is supported  in part by the Russian Foundation for
Basic Research, Grant  09-02-01149  and by the Heisenberg-Landau
program.

\end{document}